\documentclass[runningheads]{eptcs}
\providecommand{\event}{QPL 2015}
\usepackage{makeidx}

\usepackage{amsfonts,amssymb,amsmath,alltt}
\usepackage {mathpartir,amssymb}
\usepackage{graphicx}
\usepackage[utf8]{inputenc}
\usepackage{url}
\usepackage[english]{babel}

\newsavebox{\savecode}
\newenvironment{ttbox}{%
\lrbox{\savecode}%
\begin{minipage}{0.97\linewidth}%
\begin{alltt}\ttbraces\small}%
{\end{alltt}\end{minipage}\endlrbox}
\newcommand{\ttprint}{\par\usebox{\savecode}\par}

\def\ttbraces{\let\.=\nobreak\chardef\{=`\{\chardef\}=`\}\chardef\|=`\\}

\newcommand{\symb}[1]{\makebox{\it #1}} 

\newcommand{\ket}[1]{|#1\rangle}

\def\titlerunning{Formalization of Quantum Protocols using Coq}
\def\authorrunning{Boender, Kamm\"uller, Nagarajan}

\begin{document}



\title{Formalization\\ \emph{of}\\ Quantum Protocols using Coq}
\author{Jaap Boender \and Florian Kamm\"uller \and Rajagopal Nagarajan
\institute{Department of Computer Science,
  School of Science and Technology, Middlesex University, London, UK
\email{\texttt{{J.Boender,F.Kammueller,R.Nagarajan@mdx.ac.uk}}}}}

%

\maketitle

\begin{abstract}
Quantum Information Processing, which is an exciting area of research at 
the intersection of physics and computer science, has great potential
for influencing the future development of information processing systems. The building of practical, general purpose Quantum Computers may be some years 
into the future.  However, Quantum Communication and Quantum Cryptography are well developed. Commercial Quantum Key Distribution systems are
easily available and several QKD networks have been built in various parts of
the world. The security of the protocols used in these implementations rely on information-theoretic proofs, which may or may not reflect actual system behaviour. Moreover, testing of
implementations cannot guarantee the absence of bugs and errors. This paper presents a novel framework for modelling and
verifying quantum protocols and their implementations using the proof assistant Coq. We provide a Coq
library for quantum bits (qubits), quantum gates, and quantum measurement.  As a step towards
verifying practical quantum communication and security protocols such as
Quantum Key Distribution, we support multiple qubits, communication and
entanglement. We illustrate these concepts by modelling the Quantum
Teleportation Protocol, which communicates the state of an unknown quantum bit
using only a classical channel.
\end{abstract}


\section{Introduction}
\label{sec:intro}
Quantum cryptography aims to overcome the limitations of classical cryptography
by providing perfect security. The Quantum Key Distribution protocol
by Bennett and Brassard (BB84) \cite{bb:84} 
showed how quantum superposition and non-clonability of quantum states
can be used to communicate a secret key, while \emph{a posteriori} verifying
on an open line whether the communication has been intercepted. 
Thus, by sacrificing a small portion of the transmitted key, it is possible to
detect eavesdropping in almost 100\% of cases, which in turn, guarantees (almost) perfect secrecy when using the transmitted
one-time key in future communications. This protocol has been implemented
in commercial products, e.g., by Id Quantique \cite{cer:13}, MagiQ \cite{mag:13}, NEC and Toshiba, amongst others,
and has been used in practical applications, e.g., the Geneva election ballot count \cite{pec:07}. Various QKD networks have been built, including the DARPA Quantum Network in Boston, the SeCoQC network around Vienna and the Tokyo QKD Network.

Physical restrictions of quantum communication, like preserving photon states over long distances, 
are gradually being resolved, for example, by quantum repeaters \cite{Riedmatten2004a} and using 
quantum teleportation, i.e., transmission of quantum bits through a classical medium. There is no doubt 
that quantum communication and quantum cryptographic protocols will become an integral part of our society's 
infrastructure.

On the theoretical side, quantum key distribution protocols such as BB84 have been proved to be 
unconditionally secure \cite{MayersD:uncsqc}. It is important to understand that this is an information-theoretic proof,
which does not necessarily guarantee that \emph{implemented systems} are unconditionally secure. That is why alternative 
approaches, such as those based on formal methods, could be useful in analysing behaviour of implemented systems.

Quantum Formal Methods has been an active research area recently and quantum communication and protocols 
have been investigated and formalised.
Formal investigations of quantum protocols \cite{gn:05,JorrandP:towqpa,YingM:algqp} use languages for distributed
communications, like CCS and the $\pi$-calculus \cite{DBLP:journals/iandc/MilnerPW92a}
or related formalisms. Such formalisations enable reasoning about the properties of the quantum communication model using techniques like 
type checking \cite{GaySJ:typtcq}, as well as bisimulation and equational reasoning 
\cite{LalireM:relqpb,DavidsonT:forvtq,FengY:bisqp} for these underlying languages.
Other approaches which are more automated and tool-based include model checking and 
equivalence checking \cite{GaySJ:qmcmcq,FengY:modcqm,GaySJ:spevqp,ArdeshirLarijaniE:equcqp,ArdeshirLarijaniE:vercqp}. 
In our approach, which we believe is  new, we directly formalise the mathematical model of quantum 
communication in constructive logic using the Coq proof assistant.
In this way, we build a mechanised and partly automated logical theory of quantum 
communication. This theory allows the implementation of quantum cryptography protocols
based on their mathematical models and the proof of their properties.
Moreover, since Coq's logic is constructive, extraction of code is possible.

The expressivity of Coq and its code extraction properties make it an interesting
tool for the creation of generic frameworks, i.e. abstract formalisations of
practical specification and verification problems that may be instantiated to 
various applications. A famous application of Coq is the formal proof of the four colour theorem by Gonthier
\cite{gon:08}. Another impressive achievement by Gonthier, together with others, is the Coq proof of the Odd Order Theorem \cite{Gonthier:macpoo}. 
Coq has also been used for the formal verification of large
pieces of software, such as a C compiler in the CompCert
project \cite{Leroy-CompCert-CACM}. Other higher-order theorem provers, such as Isabelle, HOL, NuPRL and PVS have been used in a variety 
of applications. Mathematicians have also been using theorem proving techniques 
to formalise and prove difficult conjectures and theorems. Vladimir Voevodsky and others are using Coq to formalise homotopy theory. 
Thomas Hales has been working on 
a project called Flyspeck, formalising
his proof of the Kepler conjecture in the theorem prover HOL Light. 

The contribution of this paper is a framework for modelling and proving quantum protocols
in Coq \cite{coq:13}.  We assume that the reader is familiar with quantum theory and computing, for details we refer to the book by Nielsen and Chuang \cite{nc:00}
or the paper by Rieffel and Polak \cite{rp:00} tailored to computer scientists. We provide a Coq library for quantum bits (qubits), quantum gates, 
and quantum measurement (Section \ref{sec:qubits}). 
An important aspect of quantum information processing is the use of entanglement of qubits and their transmission.
Our Coq library also provides this, through measurement of qubit vectors (Section \ref{sec:measure}). Finally, we illustrate the
functioning and application of entanglement and measurement by modelling quantum teleportation (Section \ref{sec:teleportation}), which is a protocol 
that allows the use of so-called Einstein-Podolsky-Rosen (EPR) pairs of maximally entangled qubits
to communicate the state of a qubit from Alice to Bob using a pair of classical bits.
This technique is crucial for realistic implementations of quantum protocols over long distances
by repeaters as mentioned earlier.


\section{Interactive Proof in Coq}
Interactive theorem proving is semi-automated proof development, usually in Higher Order Logic (HOL)
or some variant, like the Calculus of Indictive Constructions that underlies 
the interactive theorem proving system Coq \cite{coq:04,coq:13}.
Coq's logic is not classical: it is a constructive type theory interpreted as a logic
according to the Curry-Howard isomorphism \cite{how:80}. 
Although constructivity imposes restrictions that can be quite awkward at times, it 
offers one decisive advantage. Since all proofs are constructions of witnesses, 
they are executable as programs. Program code can be extracted automatically
from the Coq tool into the programming language OCaml 
which is also the implementation language of the Coq system. 

Another advantage of Coq is that its type system is more powerful
than the simple type theory that is the basis for classical HOL.  For example,
the type system of
Coq allows \emph{dependent types}, which can be used to represent universal
and existential quantification. However, the type system is sufficiently restricted 
that it remains decidable.
Note, however, that there are some constructions in
Coq that violate decidability of type checking. Generally, type checking in
Coq is decidable, but, as an example, pattern matching constructions over
dependent types can become undecidable. Coq also offers dependent records and 
additionally a separate module system \cite{chr:03}. 

Coq's original foundation, the Calculus of Inductive Constructions, has been
extended to
a calculus of {\it inductive} constructions \cite{cp:90} that includes inductive
definitions. Similar to a datatype definition in a programming language like ML, 
an inductive definition in Coq consists of a set of rules describing the signature
of the constructors of the type. However, Coq's logic, according to the Curry-Howard
isomorphism, is defined by its types. Therefore, an inductive definition may also be used
to define logical formulas. 
Using inductive definitions for the formalisation of
computer science related subjects is very natural, because types defined by an
inductive definition automatically contain an induction principle and so-called
exhaustion properties that enable to reason by {\it inversion}. This means that, if we need to show
a property for all elements of a type, it suffices to make a case analysis over all
different manifestations of elements of the type given by the constructors of the
inductive definition.

An excellent introduction to 
Coq is \cite{chlipalacpdt2011}. 
More concrete features of the Coq system will be explained when we use them 
in the following sections.

Our quantum formalisation is work in progress and the current version is available online \cite{qoq:14}. 
It comprises around 2000 lines of Coq code. About 700 of these lines make up the
specification of some elementary matrix theory (there are existing matrix
libraries in Coq, but these do not work together well with the C-CoRN library).
The development is discussed in the rest of the paper.

\section{Qubits and Quantum Gates in Coq}
\label{sec:qubits}
For the formalisation of qubits and their basic infrastructure in Coq, we need theory libraries
supporting vector spaces and matrix operations. 
The constructive Coq repository C-CoRN at Nijmegen \cite{DBLP:conf/mkm/Cruz-FilipeGW04} 
offers a mathematical library that we use as a starting point for the theory of
real and complex numbers. 
We give here just a short overview of the matrix operations defined in
our framework. Further information can be found in the attached online resources.
Based on the C-CoRN module \texttt{CRings} for constructive rings, we have built a
module \texttt{MatrixOps} specifying the signature of $m \times n$-matrices and their operations over a ring.
As operations, we define matrix equality using the notation \texttt{\{=\}}, 
matrix addition \texttt{\{+\}}, matrix multiplication \texttt{\{*\}}, 
matrix scalar multiplication \texttt{\{**\}}.
We also define tensor multiplication \texttt{\{o\}} needed for multiple qubits (see Section \ref{sec:measure}). 
We specify the corresponding properties of these operations in module \texttt{MatrixSpec}.
As a first consistency check of our specification, we can already prove some algebraic properties of matrices.
Examples of such properties include facts such as matrices with addition form 
an abelian (commutative) group.
These proofs are contained in the module \texttt{MatrixSetoid}, which is used in the main formalisation 
in the file \texttt{Quantum.v}, and will be further explained below.

Given this foundation, we can define qubits as complex column vectors. 

\begin{ttbox}
Definition qubit (n: nat) :=
 \{q: matrix (2^n) 1 | vector_length q [=] [1]\}
\end{ttbox}
\begin{center}
\fbox{\ttprint}
\end{center}

This notation
defines \texttt{qubit} as a $2^n \times 1$-matrix
over the complex numbers \texttt{CC}, such that the vector length is 1
(i.e. the vector is a unit vector). In this section,
we only consider single qubits, i.e., \texttt{n} is 1.
We provide basic infrastructures for \texttt{qubits}, for example,
an equivalence relation and related proofs and auxiliary lemmas. In order to
create an element of the \texttt{qubit n} type, we will need to provide both
a matrix and a proof that the vector length of this matrix is 1.

We use our own matrix library for matrix operations. We have a type constructor
that, given two natural numbers, constructs a new Coq type representing a
matrix:

\begin{ttbox}
Parameter matrix: nat -> nat -> Type.
\end{ttbox}
\begin{center}
\fbox{\ttprint}
\end{center}

The basis vectors can be defined for any number of qubits as follows.
The following function \texttt{basis} is auxiliary.

\begin{ttbox}
Definition basis n k: (matrix n 1) :=
 create n 1 (fun i j =>
  if eq_nat_dec (`i) (S k)
  then [1]
  else [0]).
\end{ttbox}
\begin{center}
\fbox{\ttprint}
\end{center}

Then we add a lemma stating that any matrix created by calling \texttt{basis} 
represents a unit vector:

\begin{ttbox}
Lemma basis_length: forall n k,
  k < 2^n -> vector_length (basis (2^n) k) [=] [1].
\end{ttbox}
\begin{center}
\fbox{\ttprint}
\end{center}

And finally we combine the function and the lemma into the definition of a qubit. We use the
\texttt{exist} function to combine the matrix created by calling \texttt{basis}
and the proof from \texttt{basis\_length} into a member of the \texttt{qubit n}
type.

\begin{ttbox}
Definition basis_q \{n\} (i: nat | i < 2^n): qubit n :=
 exist _ (basis (2^n) (`i))
  (basis_length (n) (`i) (proj2_sig i)).
\end{ttbox}
\begin{center}
\fbox{\ttprint}
\end{center}

Note that, \texttt{basis\_q 2 0 = [1; 0; 0; 0]} and \texttt{basis\_q 2 1 = [0; 1; 0; 0]}, etc.
This is not a bug, but a feature, since the superposition states of qubits necessitate higher dimensional vector spaces, as we will see in Section \ref{sec:measure}.

In this development, we are using the standard basis for qubits. Of course, 
other bases can be used just as easily, because measurement with respect to another basis is equivalent to applying a suitable transformation and 
then measuring with respect to the standard basis. 

Quantum gates over $n$ qubits are now defined as the type of $2^n \times 2^n$ matrices.

\begin{ttbox}
Definition gate (n: nat) :=
 \{g:matrix (2^n) (2^n)|unitary g\}.
\end{ttbox}
\begin{center}
\fbox{\ttprint}
\end{center}

Our matrix library is capable of using any ring as a coefficient algebra. 
The type \texttt{gate} is a definition in the module \texttt{Quantum} representing
the specific ``subtype'' of quadratic matrices over complex numbers.
This is achieved by the trailer in \texttt{Quantum} instantiating our module
\texttt{MatrixSetoid} with C-CoRN's complex numbers \texttt{CC} 
as the base ring \texttt{C} and opening it with \texttt{Import} in the current 
context of the module \texttt{Quantum}.

\begin{ttbox}
Declare Module Matrix: MatrixSetoid with
  Definition C := (cf_crr CC).
Import Matrix.
\end{ttbox}
\begin{center}
\fbox{\ttprint}
\end{center}

We also use a dependent type to specify that gates are unitary matrices.

With this infrastructure at hand, we can define quantum gates in Coq. For example, the $X$ gate is defined as follows.

\begin{ttbox}
Definition x_function
 (i: nat | 0 < i <= 2) (j: nat | 0 < j <= 2): C :=
 match `i, `j with
 | 1, 2 | 2, 1 => [1]
 | _, _        => [0]
 end.

Program Definition x_gate: gate 1 :=
 exist (fun x => unitary x) (create 2 2 x_function) _.
\end{ttbox}
\begin{center}
\fbox{\ttprint}
\end{center}

We first define a function that matches the specific matrix coordinates with
the value at that coordinate, and use this function to create a matrix.
The \texttt{create} function works in such a way that if $M$ is the result
of calling \texttt{create} with function \texttt{f}, then
$M_{x,y} = \mathtt{f}\ x\ y$ (for all $x$ and $y$ that are valid coordinates
for the matrix $M$).

We use Coq's {\sf Program} extension to make working with dependent types
easier. It will generate the proof obligations that are needed and try to solve
them automatically. Any obligations that cannot be solved are left to the user
to prove. In this case, there is only one proof obligation: we need to prove
that the matrix is unitary. This is easy to do, since the matrix is only 
2 $\times$ 2 in size.

A gate represents a qubit transformation function. In order to apply
a gate to a qubit we define the following operator.

\begin{ttbox}
Program Definition apply \{n\} (q: qubit n) (g: gate n):
 qubit n :=
  exist (fun x => vector_length x [=] [1])
   ((`g) \{*\} (`q)) _.
\end{ttbox}
\begin{center}
\fbox{\ttprint}
\end{center}

This function multiplies the qubit and gate matrices. It needs some more
syntax because of the dependent types: first we discard the proofs from the
$g$ and $q$ parameters (by using the backquote operator) so that we keep only
the actual matrices. These we multiply, and then we use \texttt{exist} to
create a new inhabitant of a dependent type. This is necessary, since the result is a qubit,
so we will need to prove that it is a unit vector. The {\tt Program} extension
again helps us to achieve this easily.

We formalise the Hadamard gate just like any of the other quantum gates as \texttt{hadamard}. In fact,
it is the same as the \texttt{x\_gate} shown above with the matrix entries \texttt{Zero, One, One, Zero}
replaced by \texttt{onestwo, onestwo, onestwo, --onestwo}. 

Here \texttt{onestwo} represents $\frac{1}{\sqrt{2}}$
and the definition in Coq is as follows.

\begin{ttbox}
Definition onestwo: CC :=
  Build_CC\_set (One [/] NRootIR.sqrt Two
  (less_leEq _ _ _ (pos_two _)) [//] stwo\_pos) ZeroR.
\end{ttbox}
\begin{center}
\fbox{\ttprint}
\end{center}

The definition of division and square root in C-CoRN may seem complex but it
integrates mathematical accuracy. The division operator
has an extra argument ({\tt stwo\_pos}) in order to ensure that the divisor is
not zero. Similarly, the square root function takes an extra argument that
ensures that we do not take the square root of a negative number.

\section{Formalising Multiple Qubits, Measurement, and Entanglement}
\label{sec:measure}
We now explain how tensor products, measurement and entanglement
are formalised in our Coq framework. 

The Coq definition for the tensor product is a straightforward type of matrices.

\begin{ttbox}
Parameter tensor_mult: forall m n p q,
 matrix m n -> matrix p q -> matrix (m*p) (n*q).
\end{ttbox}
\begin{center}
\fbox{\ttprint}
\end{center}

The specification of tensor multiplication is the following.

\begin{equation}
\label{eqn:tensor}
M_{i,j}N_{k,l} = R_{p(i-1)+k,q(j-1)+l}
\end{equation}

When we formalise this definition in Coq, we need to provide the proofs that the indices 
are within bounds
(for example, that $0 < p(i-1)+ k \leq m*p$). To state the fact that two
indices that are within the source matrices are also within bounds in the result
matrix, we use the lemma
{\tt tensor\_bounds} (the backquote is used to extract the number from the
dependent type):

\begin{ttbox}
Lemma tensor_bounds: forall m p
 (i: nat | 0 < i <= m) (k: nat | 0 < k <= p),
  0 < p*(`i-1)+(`k) <= m*p.
\end{ttbox}
\begin{center}
\fbox{\ttprint}
\end{center}

This lemma can then be used in the specification of tensor product multiplication. To 
improve the readability of formulas we define the infix notation 
\texttt{\texttt{\{o\}}} for \texttt{tensor\_mult}. The definition then maps the 
tensor multiplication to the complex number multiplication \texttt{[*]}
following the indexing of definition (\ref{eqn:tensor}) while supplying 
proofs that indices are in range.

\begin{ttbox}
Parameter tensor_mult_spec1:
 forall m n p q (mx1: matrix m n) (mx2: matrix p q)
 i j k l,
  (mx1 \{o\} mx2) \{exist (fun x => 0 < x <= m*p)
                  (p*(`i-1)+`k)
                 (tensor\_bounds m p i k),
                 exist (fun x => 0 < x <= n*q)
                  (q*(`j-1)+`l)
                 (tensor\_bounds n q j l)\}
 [=]
 mx1 \{i, j\} [*] mx2 \{k, l\}.
\end{ttbox}
\begin{center}
\fbox{\ttprint}
\end{center}

In order to represent the use of the tensor product to combine gates and
qubits, we prove the following theorem (in {\tt MatrixTheory}).

\begin{ttbox}
Theorem mult_dist_tensor: forall \{m1 n1 p1 m2 n2 p2\}
 (mx1: matrix m1 n1) (mx2: matrix n1 p1)
 (mx3: matrix m2 n2) (mx4: matrix n2 p2),
 (mx1 \{*\} mx2) \{o\} (mx3 \{*\} mx4) \{=\}
 (mx1 \{o\} mx3) \{*\} (mx2 \{o\} mx4).
\end{ttbox}
\begin{center}
\fbox{\ttprint}
\end{center}

\subsection{Measurement}
For modelling purposes, we can treat measurement of multiple qubits as
sequential measurement of single qubits.

In our model, measurement of a quantum state consists of taking the state 
(represented by the \texttt{qubit n} Coq type) and a bit number, and
then computing the probabilities of measuring a 0 or a 1 for that bit.

As a quantum state is represented by a matrix, and each element of this matrix
represents one possible measurement result, measuring one specific bit amounts 
to taking the sum of the relevant elements.

For example, in a three-qubit state, there are eight elements, of which the first
represents the probability of measuring 000, the second the probability of
measuring 001, and so on. If we want to measure the second bit, the probability
of measuring 0 for this bit is equal to the sum of the probabilities of
measuring 000, 001, 100 and 101.

The \texttt{sum\_pair} function takes a bit number $i$ and a quantum state $q$.
It returns a pair of probabilities for bit $i$: that of measuring 0 and that of measuring 1.

\begin{ttbox}
Definition sum_pair {n} i (q: qubit n): IR * IR :=
 foldn (2^n) (2^n) (le_refl (2^n)) (fun k Hk acc =>
  match acc with
  | (acc1, acc2) =>
    if digit_is_zero i k
    then (acc1 [+]
     AbsCC ((`q) \{exist _ _ (plusone _ _ Hk),
      exist _ _ (plusone _ _ (lt_0_Sn 0))\}) [^] 2,
     acc2)
    else (acc1, acc2 [+]
     AbsCC ((`q) \{exist _ _ (plusone _ _ Hk),
      exist _ _ (plusone _ _ (lt_0_Sn 0))\}) [^] 2)
  end
 ) ([0], [0]).
\end{ttbox}
\begin{center}
\fbox{\ttprint}
\end{center}

In order to compute the new quantum state after a measurement, half of the
values in the matrix will become 0 (if we have just measured that bit 5 is
0, then all the states that represent bit 5 being 1 have become impossible!).
However, a quantum state is a unit vector, so we will have to normalise the 
remaining values to keep this property. We do this by dividing each value by
the sum of all remaining values. In this way, the probability distribution for
the bits that have not been measured remains the same.

We need to be sure that this is not a division by zero. Therefore we introduce
the following axiom, saying that it is decidable whether the probability of
measuring 0 or 1 for a given bit is 0 or not. This is not very farfetched,
since an event that has probability 0 will never happen. These axioms
internalise that knowledge.

\begin{ttbox}
Axiom sum_pair1: forall \{n\} (i: nat | i < n)
 (q: qubit n),
  fst (sum_pair (`i) q) [=] [0] or
  [0] [<] fst (sum_pair (`i) q).

Axiom sum_pair2: forall \{n\} (i: nat | i < n)
 (q: qubit n),
  snd (sum_pair (`i) q) [=] [0] or
  [0] [<] snd (sum_pair (`i) q).
\end{ttbox}
\begin{center}
\fbox{\ttprint}
\end{center}

Now we can define the function {\tt nqv} (for `new qubit vector') that 
computes the new quantum state after a measurement. The parameter $f$ should
be either the identity function or the boolean NOT function. This allows us
to compute both the new quantum state after a measurement of 0 and after a 
measurement of 1 with the same function. We also add the sum, or the
probability of measuring the result, and a proof that it is greater than 0,
needed for the division.

\begin{ttbox}
Definition nqv \{n\} (k: nat) (f: bool -> bool)
 (sum: IR) (sum_proof: [0] [<] sum) (q: qubit n) :=
 create (2^n) 1 (fun i j => 
  if f (digit_is_zero k (`i))
  then (`q) {i, j} [/]
   cc_IR (NRootIR.sqrt sum (less_leEq _ _ _ sum_proof))
   [//] (re_ap_zero _ (ap_symmetric _ _ _
    (less_imp_ap _ _ _ (NRoot_pos _ _ _ _ sum_proof))))
  else [0]
 ).
\end{ttbox}
\begin{center}
\fbox{\ttprint}
\end{center}

The division needs three arguments: two for the division operation and one for 
a proof that the divisor is not 0. 

Note that these functions are all auxiliary 
and not intended to be called on
their own. The main function that takes care of measurement is shown below.

\begin{ttbox}
\small
Program Definition measure \{n\} (i: nat | i < n)
 (q: qubit n): list (IR * qubit n) :=
 match sum_pair1 i q with
 | inl _ => (* zero *) [([1], q)]%list
 | inr sum0_gt => 
   match sum_pair2 i q with
   | inl _ => (* zero *) [([1], q)]%list
   | inr sum1_gt =>
     [(fst (sum_pair i q),
      existT _ (nqv (`i) negb (fst (sum_pair i q))
       sum0_gt q) _);
      (snd (sum_pair i q),
      existT _ (nqv (`i) (fun x => x)
       (snd (sum_pair i q)) sum1_gt q) _)]%list
   end
 end.
\end{ttbox}
\begin{center}
\fbox{\ttprint}
\end{center}

This function uses the {\tt sum\_pair} axioms and the {\tt nqv} function.
It takes the two probabilities (of measuring 0 and 1 for bit i) and computes
the new quantum states that occur after this measurement. If either
probability is zero, we cannot compute the new state, because this would involve
a divition by zero. However, the fact that the probability is zero means that
this situation will never occur. The {\tt sum\_pair} axioms are a way of
externalising that knowledge.

\subsection{Entanglement and EPR pairs}

Measurement also allows us to think about entanglement. As mentioned 
earlier, definition of an entangled state is a state that cannot be broken
down as the tensor product of smaller states. However, in constructive logics
such as the one used by Coq, it is difficult to
prove the absence of something---it would require proving that any breakdown
in some way leads to a contradiction. Fortunately, we can use an alternative definition 
of entanglement which is specified in terms of measurement. Two qubits in a quantum state are
entangled if measuring one qubit changes the state of the other; a quantum state 
space is entangled if it contains entangled qubits.

The advantage of this definition is that it is much easier to work with: we
just need to show that two qubits are entangled. In Coq, this definition becomes:
\newcommand{\BACKSLASH}{\char`\\}

\begin{ttbox}
Definition entangled_p {n} (q: qubit n)
 (p1: nat | p1 < n) (p2: nat | p2 < n) :=
  exists pr, exists res,
    List.In (pr, res) (measure p1 q) /\symbol{92}
         ~(distribution_equal (measure p2 res) (measure p2 q)).
\end{ttbox}
\begin{center}
\fbox{\ttprint}
\end{center}

This then allows us to define entanglement: a state is entangled if it
contains entangled qubits:

\begin{ttbox}
Definition entangled {n} (q: qubit n) :=
  exists p1 p2, (`p1) <> (`p2) /\symbol{92} entangled_p q p1 p2.
\end{ttbox}
\begin{center}
\fbox{\ttprint}
\end{center}

The {\tt distribution\_equal} function determines whether two probability
distributions are equal.

The {\tt entangled\_p} definition, therefore, says that the probability of $p2$
being $v$ in $q$ must not be the same as the probability of $p2$ being $v$ in
$res$, where $res$ is the result of measuring $p1$ in $q$.

As a sanity check, we can now prove that the `maximally entangled' EPR pair
$\frac{1}{\sqrt{2}}(\ket{00}+\ket{11})$ is indeed entangled:

\begin{ttbox}
Definition epr_function (i: nat | 0 < i <= 4)
 (j: nat | 0 < j <= 1): C :=
  match `i with
  | 1 | 4 => onestwo
  | _     => [0]
  end.

Program Definition epr_1: qubit 2 :=
  (exist (fun x => vector_length x [=] [1])
  (create _ _ epr_function) _).

Lemma entangled: entangled epr_1.
\end{ttbox}
\begin{center}
\fbox{\ttprint}
\end{center}

Note how we specify a qubit: we simply give a vector with 4 elements and then
explicitly note that this is an element of {\tt qubit 2}. Coq automatically
checks that the types are equivalent.

We can also prove that the state $\frac{1}{\sqrt{2}}(\ket{0}+\ket{1}) \otimes \frac{1}{\sqrt{2}}(\ket{0}+\ket{1})$ is not entangled:

\begin{ttbox}
Lemma entangled2: ~entangled ([cc_IR Half;
  cc_IR Half; cc_IR Half; cc_IR Half]: qubit 2).
\end{ttbox}
\begin{center}
\fbox{\ttprint}
\end{center}

\section{Quantum Teleportation in Coq}
\label{sec:teleportation}
The quantum teleportation procedure \cite{bbcjp:93} allows the communication of the state of an unknown qubit, 
from one party to another, via a classical (\emph{i.e.,} non-quantum) medium.
According to the procedure, Alice has a qubit $\phi = a\ket{0} + b\ket{1}$ whose state she does not  
know and which she wishes to send to Bob.
Alice and Bob each control one qubit of the maximally entangled EPR pair 
$\psi = \frac{1}{\sqrt{2}}(\ket{00} + \ket{11})$, i.e., we assume that the entangled EPR pair 
$\psi$ is shared between Alice and Bob.
The translation of this protocol into our Coq framework is relatively straightforward.
We can use the definition of the maximally entangled EPR pair $\psi$ 
from the previous section ({\tt epr\_1}).
Alice's function is a sequence of transformations and measurements applied to the qubit-vector
composed of $\phi$ and $\psi$ \cite{rp:00} leaving Bob's third qubit unchanged ($I$-gate).
\[
\begin{array}{ccc}
{\rm Alice}_f & = & (H \otimes I \otimes I)(C_{\symb{not}} \otimes I)(\phi \otimes \psi)\,.
\end{array}
\]
Alice then measures the first two qubits, thereby affecting Bob's qubit via entanglement, 
and sends two classical bits $x, y$ encoding the outcome of this measurement to Bob.
Bob can now restore $\phi$ in his second qubit of the EPR-pair by simply applying the
$I, X, Z$ or $Y$ gate to it depending on $x$ and $y$. The details of the mathematical
description \cite{rp:00} can be used one to one in the Coq definitions of teleportation below.

To compose the application of quantum gates, Alice employs the \texttt{\{o\}} operator 
which represents the tensor product. She needs the \texttt{c\_not\_gate} and the 
Hadamard transformations composed with identity transformations. 
\begin{ttbox}
Definition firstgate: gate 3 :=  c_not_gate \{o\} identity.
Definition sndgate: gate 3 :=  hadamard \{o\} identity \{o\} identity.
\end{ttbox}
\begin{center}
\fbox{\ttprint}
\end{center}
The function \texttt{Alice} applies first the two gates \texttt{firstgate} and
\texttt{secondgate} to the qubit triple \texttt{phi \{o\} epr\_1} as defined
in Alice$_f$ above. The first two qubits \texttt{p1} and \texttt{p2} of 
this application are then measured using the constructor \texttt{measure} 
(see Section \ref{sec:measure}). The result of function Alice lists the 
different outcomes of the measurement with their probabilities attached.
\begin{ttbox}
Definition Alice (phi: qubit 1): list (IR * qubit 3):= 
measure p1 (measure p2 (apply (apply (phi \{o\} epr_1) firstgate) sndgate)).
\end{ttbox}
\begin{center}
\fbox{\ttprint}
\end{center}
An application of function \texttt{Alice} to a single qubit \texttt{phi}
results in the following four possibilities;
\texttt{alpha phi} and \texttt{beta phi} are the components $a, b$, respectively, 
of qubit $\phi = a\ket{0} + b\ket{1}$.
\begin{ttbox}
Definition Alice_pos (phi: qubit 1)(q: qubit 3):=
`q = [alpha phi;beta phi;[0];[0];[0];[0];[0];[0]] 
or
`q = [[0];[0];beta phi;alpha phi;[0];[0];[0];[0]] 
or
`q = [[0];[0];[0];[0];alpha phi;[--](beta phi);[0];[0]] 
or
`q = [[0];[0];[0];[0];[0];[0];[--](beta phi);alpha phi]
\end{ttbox}
\begin{center}
\fbox{\ttprint}
\end{center}
The fact that the application of Alice's function yields precisely those
four possibilities is encoded in the following theorem. 
\begin{ttbox}
Theorem Alice_case: forall phi: qubit 1, forall q: qubit 3, forall r: IR, 
                    List.In (r,q)(`(Alice phi)) -> Alice_pos phi q.
\end{ttbox}
\begin{center}
\fbox{\ttprint}
\end{center}
For the transmission of the measurement of Alice's first two qubits we need
classical bits encoded as an inductive type to enable subsequent pattern matching.
\begin{ttbox}
Inductive Bit : Set := | z | o. 
\end{ttbox}
\begin{center}
\fbox{\ttprint}
\end{center}
Depending on the outcome of the measurement according to \texttt{Alice\_pos}, 
we define the following function that encodes this measurement in a pair of
bits. The constructors \texttt{inl} and \texttt{inr} allow
to pattern match a disjunction in Coq.
\begin{ttbox}
Definition Alice_out(phi: qubit 1)(q: qubit 3)(qp: Alice_pos phi q): (Bit * Bit) := 
  match qp with
  | inl _ => (z,z)
  | inr(inl _ ) => (z,o) 
  | inr(inr(inl  _)) => (o,z)
  | inr(inr(inr _)) => (o,o)
  end.
\end{ttbox}
\begin{center}
\fbox{\ttprint}
\end{center}
Bob operates on the transformed qubit triple \texttt{psix}, i.e. $\phi \otimes \psi$
with the first two bits collapsed and the third one---the one he controls---in an unknown 
state. But, using Coq's pattern matching, Bob can restore the original state of $\phi$ in his 
third qubit from the two classical bits he receives.
\begin{ttbox}
Definition Bob (pxy: qubit 3 * (Bit* Bit)): qubit 3 := 
  match pxy with
  | (psix,(z,z)) => apply psix (exist \_ (`identity \{o\} `identity \{o\} `identity) \_)
  | (psix,(z,o)) => apply psix (exist \_ (`identity \{o\} `identity \{o\} `x_gate) \_)
  | (psix,(o,z)) => apply psix (exist \_ (`identity \{o\} `identity \{o\} `z_gate) \_)
  | (psix,(o,o)) => apply psix (exist \_ (`identity \{o\} `identity \{o\} `y_gate) \_)
  end.
\end{ttbox}
\begin{center}
\fbox{\ttprint}
\end{center}
Now, to formalise that the protocol does actually achieve the teleportation
of Alice's qubit $\phi$, we can simply state that the combination of Alice's and Bob's
functions results in a triple of qubits whose last element is the same as $\phi$. 
Since Alice's function used $\phi$ as its first qubit, this statement then encodes that
$\phi$ has been ``teleported'' from the first position to the third position, i.e., from
Alice to Bob.
The protocol composes Alice's function, given by the theorem \texttt{Alice\_case},
followed by sending the classical bits using \texttt{Alice\_out}, and finally
applying Bob's decoding function. This combination can be applied to any of 
the possible qubits \texttt{q} in the result list \texttt{qp} that is returned
by \texttt{Alice phi}.
The first two elements measured and encoded in classical bits determine
the qubit \texttt{q}.
We represent these first two measured bits by an existentially quantified qubit pair 
\texttt{z}. The proof of teleportation consists of showing that, for each of the four possible
outcomes for \texttt{q}, a suitable \texttt{z} exists.
\begin{ttbox}
Theorem teleportation: 
forall phi: qubit 1, forall q: qubit 3,
forall pr: IR, forall qp: List.In (pr, q)`(Alice phi),
 exists z: qubit 2, 
   `Bob(q, Alice_out phi q (Alice_case phi q pr qp)) \{=\} `(z \{o\} phi).
\end{ttbox}
\begin{center}
\fbox{\ttprint}
\end{center}

\section{Conclusions and Future Work}
We have presented a framework for modelling and analysing quantum protocols using the proof assistant Coq. The framework makes it possible to represent and reason about all the essential features of quantum systems such as single and multiple qubits, quantum gates, measurement and entanglement. We have modelled and analysed the canonical example of Quantum Teleportation. As mentioned earlier, 
model-checking and equivalence checking techniques have been used for verification of quantum protocols. The work in this paper also opens up the possibility of 
combining theorem proving with these (more automatic) approaches to greater gain. 

The framework uses existing Coq libraries as much as possible, though a new 
matrix library had to be developed. This library can also be used
independently. Both the library and the framework make use of dependent types.
This can make the proofs more complicated, but it allows for more concise
formulations of key lemmas. We intend to push the use of dependent types still
further during future development of the framework.


Future work will aim to look at simplifying some definitions, improving the specifications and enhancing the general readability of the Coq code. 
We will also investigate using linear typing as a way to incorporate the notion of non-cloneability of quantum states in our framework. Our Coq development 
is already quite substantial and contains most of the necessary features to analyse large scale examples. In the near future, we hope to apply it to various 
case studies ranging from Quantum Bit Commitment and Blind Quantum Computing to Quantum Error Correction Protocols. Of course, an ambitious and 
challenging project which we have in mind is to formalise and prove the security of Quantum Key Distribution using Coq. 

\section*{Acknowledgements} We would like to thank Andrei Popescu for reading the paper carefully and providing valuable comments.

\bibliographystyle{eptcs}
\bibliography{biblio}


\end{document}